%% file: main.tex
\documentclass[letterpaper]{article} 
\usepackage{aaai2026}  
\usepackage{times}  
\usepackage{helvet}  
\usepackage{courier}  
\usepackage[hyphens]{url}  
\usepackage{graphicx} 
\usepackage{natbib}  
\usepackage{caption} 

\usepackage{shortcuts}

\usepackage{algorithm}
\usepackage{algorithmic}
\usepackage{dsfont}
\usepackage{amsmath}

\usepackage[export]{adjustbox}
\usepackage{booktabs, threeparttable, dcolumn, multirow}
\usepackage{tabularx}
\usepackage{url}
\usepackage{xcolor}
\usepackage{graphicx}
\usepackage{subcaption}
\usepackage{lipsum}
\usepackage{mdframed}
\usepackage{newfloat}
\usepackage{listings}
\DeclareCaptionStyle{ruled}{labelfont=normalfont,labelsep=colon,strut=off} 
\floatstyle{ruled}
\newfloat{listing}{tb}{lst}{}
\floatname{listing}{Listing}
\newmdtheoremenv{theo}{Definition}

\title{Anti-Establishment Sentiment on TikTok: \\ Implications for Understanding Influence(rs) and Expertise on Social Media}

\author {
    Tianliang Xu\textsuperscript{\rm 1},
    Ariel Hasell\textsuperscript{\rm 2},
    Sabina Tomkins\textsuperscript{\rm 1}
}
\affiliations {
    \textsuperscript{\rm 1} University of Michigan School of Information\\
    \textsuperscript{\rm 2}University of Michigan Department of Communication and Media\\
    tianlix@umich.edu, 
    hasell@umich.edu, 
    stomkins@umich.edu
}

\usepackage{bibentry}
\begin{document}

\maketitle

\begin{abstract}
Distrust of public serving institutions and anti-establishment views are on the rise (especially in the U.S.). As people turn to social media for information, it is imperative to understand whether and how social media environments 
may be contributing to distrust of institutions. In social media, content creators, influencers, and other opinion leaders often position themselves as having expertise and authority on a range of topics from health to politics, and in many cases devalue and dismiss institutional expertise to build a following and increase their own visibility. However, the extent to which this content appears and whether such content increases engagement is unclear. This study analyzes the prevalence of anti-establishment sentiment (AES) on the social media platform TikTok. Despite its popularity as a source of information, TikTok remains relatively understudied and may provide important insights into how people form  attitudes towards institutions. We employ a computational approach to label TikTok posts 
as containing AES or not across topical domains where content creators tend to frame themselves as experts: 
finance and wellness. As a comparison, we also consider the topic 
of conspiracy theories, where AES is expected to be common. 
We find that AES is most prevalent in conspiracy theory content, 
and relatively rare in content related to the other two topics. However, we find that engagement patterns with such content vary by area, and that there may be platform incentives for users to post content that expresses anti-establishment sentiment. 
\end{abstract}

\begin{links}
    \link{Code}{https://github.com/politechlab/AES}
\end{links}

\section{Introduction}
In recent years, growing support for populism and anti-intellectualism have drawn increased scholarly attention to the role anti-establishment views play in undermining the health of democratic societies \citep{droste2021feeling, merkley2020anti, oliver2016rise}.

In the U.S., citizens across the political spectrum increasingly distrust public serving institutions like government, the news media, scientists, and universities.   Many have developed views that are hostile towards those institutions, with consequences ranging from increased misinformation to political violence \citep{armaly2024supports, enders2023modeling, enders2021misinformation, lyons2023orientations, stecula2020trust}.
To understand and address large scale, global problems like public health crises, climate change, and democratic backsliding, there is a need to better understand the factors that might influence public opinion regarding public serving institutions and democratic systems. 

Anti-establishment views include an array of attitudes including populism, anti-intellectualism, and conspiratorial thinking, among others \citep{enders2021misinformation, oliver2016rise, uscinski2021american}, and can broadly be defined as ``the politics of opposition to those wielding power''\citep[page 31]{barr2009populists}. Such views position the ``virtuous'' ordinary citizens against the ``corrupt'' and ``immoral'' elite establishment, and generally the establishment can refer to any cultural, political, or economic elite \citep{droste2021feeling, mudde2004populist}.

Importantly, the perception of who or what is part of the elite establishment can vary over time or across individuals, but these views consistently position everyday or ordinary people as morally better than experts or those working within elite institutions. For example, while trust in science is generally high in the U.S., there is increasing sentiment that scientists and universities are a part of a powerful, elite group and are attempting to be arbiters of ``truth'' while dismissing and undermining ``common sense'', in an effort to take power away from ordinary people \citep{mede2020science}. Anti-establishment views can correlate with political ideology in certain contexts, but research shows that anti-establishment views function orthogonally to the traditional liberal-conservative political spectrum in the U.S. \citep[e.g.,][]{uscinski2021american}. The extent to which Americans hold anti-establishment views has varied over time, and while the majority of citizens do not hold such views, the prevalence of these views is increasing \citep{droste2021feeling, mede2020science, merkley2020anti}.
    
Most research in this area has focused on anti-establishment views aimed at the political and economic elite, like populism \citep[e.g.,][]{uscinski2021american}, but there has been increased attention towards views that are hostile towards other perceived elites, like scientists, experts, and doctors \citep{dyor_ariel, eberl2021populism, merkley2020anti, mede2020science, oliver2016rise}. 
Though there are many reasons why people may come to hold anti-establishment views, from personality traits, to economic resentment and rising political corruption \citep{piazza2024populism, uscinski2021american}. Researchers have begun to explore the relationship between anti-establishment views and social media use, including both political and apolitical social media \citep{dyor_ariel, uscinski2021american}.
This research suggests that individuals with anti-establishment views may be drawn to extremist social media content that promotes anti-establishment content, like conspiracy theories \citep{uscinski2021american,mitra2021understanding}, but also that the incentive structures of social media platforms may cultivate and reward content that promotes anti-establishment views \citep{chinn2026does, hasell2023political, starbird2023influence, tripodi2024your}.
There is some evidence of research conducted by platforms like Facebook and Twitter themselves finding that their platforms tend to reward and amplify political content that promotes anti-establishment views \citep{milmo2021twitter}. However, there is little research that systematically examines how much content posted in social media may express anti-establishment views in seemingly apolitical contexts. 

AES: We operationalize anti-establishment sentiment (AES) as expressions of distrust or criticism of public serving institutions for purposefully working against the interests of the American people (the People) \citep{droste2021feeling, mudde2004populist}. We then focus on examining when, how, and how often AES occurs on social media. To be precise, we address the following research questions: 

\begin{itemize}
  \item What is the relationship between post topic and AES content?
    \begin{itemize}
    \item How does the use of AES differ by post topic on TikTok?
    \item How do users engage with content with AES and without it and how does that engagement vary across topics? 
    \item Which institutions are most likely to be the target of AES across topics?
    \item Is the relationship between post topic and AES content similar across social media platforms?
    \end{itemize} 
    \item How prevalent is AES content on the For You Page?
    \item How do videos with AES content use linguistic style to create divisions between the People and elites?
\end{itemize}

Our first set of questions concerns how users interested in different topics may encounter AES in social media. 
AES research typically focuses on a given community, from healthcare \citep{chinn2026does}, 
to conspiracy \citep{samory2018government}. We would like to systematically understand how the topics one is interested in influences
their exposure to AES on TikTok, as well as how content creators in different genres are incentivized to create (or not) AES content, and 
the kinds of institutions they target.
In order to address these questions we collect a core dataset of 26,783 videos using the TikTok Research API. 
We then develop a training program to annotate the data with the help of crowd-sourced labelers. 
Finally, we train supervised learning models to annotate the remaining unlabeled data. 
Using the final human+machine labeled data, 
we address the questions above which depend on the relationship between post topic and AES content on TikTok. 

Additionally, to understand whether the relationship between topic and AES prevalence generalizes to other platforms, we collect an additional dataset from YouTube. We sample and manually label a select number of posts from the three topics of conspiracy, finance, and wellness collected with the YouTube API.
Finally, in order to understand how a user who is not systematically interested in a single topic may encounter AES content
we collect data from the TikTok For You Page (FYP).
To do so, we employ sock puppets to simulate interactions with the platform.

Finally, we are interested in how content creators on TikTok use linguistic style to create divisions between the People and elites. 
Such divisions are generally understood to be an important component of AES messaging \citep{droste2021feeling,mudde2004populist}. 
Yet, it is not understood how this messaging might appear on TikTok, a platform which has been found to have a unique style \citep{herrman2019tiktok}.

\section{Related Work}
Despite increased interest in the role of anti-establishment views in democratic society, and the connection between those views and social media use, there has been little systematic analysis of the prevalence of AES in social media content, especially outside of political contexts or conspiracy theories. Research in the humanities and social sciences have shown that AES often appears in social media content. In some cases, this is because strategic, populist political actors are encouraging distrust and hostility towards established political actors and news media for their own purposes, and social media provides the affordances and capabilities for political influence outside of institutions \citep{tripodi2022propagandists, tripodi2024your}. In other cases, AES in social media content may be the result of everyday users navigating a crowded and cacophonous information environment at a time when most people are skeptical and distrustful of established, mainstream information sources \citep{chinn2026does, cotter2022judging, Knight_Foundation}.
As content creators compete for attention and influence, they are incentivized to establish themselves as credible sources of information, often positioning themselves as aligned with the people and in opposition to established institutions and industries \citep{chinn2026does, hasell2023political}. However, the extent to which content may actually reflect AES remains unclear. 

Within the computational social science community there has been work on detecting conspiracy theories \citep{diab2024classifying}, 
on 
understanding conspiracy talk as collective sensemaking \citep{kou2017conspiracy}, 
and on detecting, describing and understanding conspiratorial language \citep{steffen2023codes,samory2018government,starbird2017examining}. 
This work tends to take conspiratorial discussions as a starting point, 
inspecting conspiracy theories in online spaces directly. 
There has also been work looking at populist rhetoric and political extremism
and how the political establishment is portrayed negatively  
within such language
\citep{grover2019detecting,jungherr2022populist}. 
However, 
anti-establishment views encapsulate far more than politics and government, 
as fears about hidden agendas of powerful actors pervade many industries, including medicine, banking, and agriculture, among others. 
For example, recent work has shown that there is a relationship between attitudes towards vaccinations and trust 
in institutions \citep{mitra2021understanding}.

Our study examines how frequently such anti-establishment sentiments appear in social media content and whether such content may be encouraged by users via social media engagement, including likes, comments, and shares. We specifically examine two topic areas,  finance and wellness, that are popular in social media generally, but are not necessarily associated with AES. We compare these to the area of conspiracy, which generally is considered to be a topic with high amounts of AES.

Finally, we analyze a sample of posts which were collected in Fall 2024 with the use of ``sock puppets'' \citep{bandy2021problematic}. These posts from the FYP give us a glimpse of a random sample of Tik Tok content and  provide an estimate of the AES content a typical user may encounter across topics. 

\section{A Computational Approach for Detecting AES}
We contribute a novel study of AES on the social media platform TikTok, which is an increasingly popular platform in the U.S., especially among young people \citep{pew2024a, pew2024b}. 
Though mostly used for entertainment, of those under 30, 50\% now use TikTok as a source of news and information and 75\% rely on TikTok for product reviews and recommendations by content creators \citep{pew2024a, pew2024b}. 
This highlights that TikTok is increasingly seen as a useful source of information on a range of topics among young people.

To examine our research questions, we first describe our conceptualization of AES. Next, we translate this into a supervised learning task. We then describe our process for obtaining human annotations of this concept. Finally, we apply machine learning to annotate the remaining data
and assess the performance of the machine learning method. 

\begin{theo}
Fundamentally, an anti-establishment view expresses that institutions are working \textit{against} the interests of the American people.
An anti-establishment view is 
any comment that expresses dislike, distrust, or criticism of a mainstream organization or industry in America related to a public serving institution (i.e., the government, NASA, the news media, universities, the EPA, etc.), health (doctors, scientists, the pharmaceutical industry, etc.), or finance (the stock market, banks, credit cards, financial industry, etc.) for \textit{working against the people}.

\end{theo}

\xhdr{Conceptualization of AES}
We  describe 
AES as shown in Definition 1. We presented this definition to 
human annotators 
when they are initially trained to perform the annotation. 
The definition we provide is more specific than typically academic definitions of anti-establishment views discussed above as we wanted to provide clear guidance to the annotators that reflect the types of messages that appear in social media. 

\subsection{Areas of focus}
We focus on three topics: conspiracy, finance, and wellness. Our first topic \textit{conspiracy}, is a natural topic to study for AES content because conspiracies often posit that small groups of powerful elites have organized events or suppressed knowledge in a way that disadvantages ordinary people \citep{douglas2019understanding}. We expect that this topic will serve as an upper bound of how much AES content we may see within a given topic. We note that not all conspiracy content is related to AES -- often discussions of conspiracies are focused on their inconsistencies, absurdities, or debunking them. In contrast, we do not expect AES to be as common in finance and wellness. However, prior research suggests that these topics will express some amount of AES.

The next topic, \textit{finance}, was selected because it is a popular topic on which people seek information and advice on social media, while related industries like the stock market, real estate, and banking are often criticized as being rigged against or unfair to ordinary people. We expect much of the finance content to be focused on utilitarian information, tips, and advice, as almost 80 percent of the Millennial and Gen Z demographic rely on social media for financial advice \citep{forbes}. However, given rising income inequality and the sense that the rich have too much political power \citep{wike2025economic}, it is also likely that some of this discussion contains AES. For example, crypto-currencies are often celebrated for operating outside of elite institutions and regulatory systems that ``favor'' the wealthy.

We examine \textit{Wellness} content as it is a multi-trillion-dollar industry that has become highly popular online \citep{callaghan2021feeling}. Much of the wellness content in social media is devoted to selling products and services in efforts to achieve better physical and mental health outcomes, but it can also contain AES \citep{chinn2023mapping}. Wellness often criticizes western medicine and the pharmaceutical industry as wanting to keep people sick to profit off disease in ways that are often anti-establishment \citep{baker2022alt}. Content creators have an incentive to promote their own expertise over the advice of doctors and other experts to build their followings, and often position intuition and common sense over medical expertise \citep{carrion2018you}.

We selected these topics as exemplars of the kinds of areas on social media where content creators build a brand around sharing advice based on their own experience, research, and expertise.   Our work provides a  reference point for future work, by  establishing a baseline of AES content in online communities with incentives to express AES or not.

\subsection{Problem Formulation}
Our goal is to solve the task of detecting the presence of AES
(labeled according to Definition 1).
Let $x$ be a social media post which can be described 
by a multimodal feature space. 
For example, $x$, may be described by textual or visual features.
We would like to determine if $x$ expresses 
AES. 

Given a dataset of $\mathds{D}= \{(x_1,y_1),\dots, (x_N,y_N)\}$
 where $y_i$ can be described by a binary label, and 
 $x_i$ belongs to a multimodal feature space $\mathcal{X}$ which depends on the available data, 
 our goal is to
 learn a function $f: \mathcal{X} \to \{0,1\}$.

\subsection{Concept elicitation with human annotation}
To detect the presence of AES, we employ human annotators. We utilized Prolific \citep{prolific2025}, a research-focused online platform, to recruit participants for the annotation task.
These annotators must first go through a training program before they are asked to perform data annotation. We conducted both the training program and the main annotation through Qualtrics \citep{qualtrics2025}, an online survey tool. 

Our survey was approved by an internal IRB. 
All annotators gave informed consent to both the training 
and the annotation task. 
The training and annotation contained warnings that annotators 
may encounter explicit content. 
Annotators were given the option to not review any content that they preferred not to review.

\xhdr{Training:} The training program includes: (1) an initial description of the task, 
(2) examples of posts which do and do not include the target concept, (3)
test questions with feedback where annotators can practice their understanding of the concepts, and (4) a final assessment. 
In the initial description of the task we explicitly define the concept of AES and provide examples of common institutions. 

We then ask them to determine if a video mentions an institution or not and provide feedback about their responses. 
In particular we ask if any of the following institutions are mentioned in the post:
\begin{itemize}
    \item The federal government or government agencies (e.g. NASA)
 \item the news media (e.g. Washington Post or Fox News)
 \item the banking industry (e.g. Wall Street)
 \item the pharmaceutical industry or medical research  (e.g. Big Pharma)
 \item  scientific research or organizations (e.g. the EPA or an Academic University)
 \item  other institution such as agriculture, the oil industry, or politicians
    
\end{itemize}

Next, we show example posts and ask participants to determine whether the post expresses AES. 
We then provide feedback on the selection. 
For example, 
consider a video with the following phrase ``If you wear a NASA t-shirt and it's not a parody. Man they got you. You're in a trance dog.''. 
We say that this post does contain AES. 
If the participant was correct we reiterate why it expresses AES (negative sentiment towards the institution NASA), 
if they are incorrect we point out what they may have missed (NASA is a federally funded entity and we would consider it an institution. The poster is generally negative about NASA). 
We also include examples of posts which do not contain AES, and again provide feedback depending 
on the participant's selection. 

After participants have interacted with examples around
whether a post mentions an institution and 
whether a post expresses AES, 
we turn to examples around comments. 
Here, we are interested in (1) whether a comment appears to agree with the post, 
(2) whether a comment expresses AES.

We describe agreement as: the comment praises the video,
expresses gratitude towards the content creator,
or criticizes those who disagree with the video. In contrast, 
we describe disagreement as: 
the comment criticizes the content creator's viewpoint,
makes fun of the video,
questions the credibility of the video,
or dismisses the content creator's values.
We follow the same format as above, except that here participants are shown both a post and a comment made in response to the post. They are then asked to determine agreement with the post and the AES content of the comment. 
They are given feedback about each example they complete. 

The final step in the training is the assessment. The assessment contains 16 questions. In order to be eligible for the annotation task, annotators must achieve a score of 75\% on the assessment.

\xhdr{Annotation:}
We structure the task around pairs of videos and comments. 
Before performing the annotation, an annotator must pass a brief assessment of 4 questions, to demonstrate that they have retained
the information from the training. 
That is, an annotator is shown a video and then shown a comment posted in response to the video. 
They are then asked the questions shown in \tabref{tab:dimensions_questions}.

\begin{table*}[h!]
\centering
\small
\renewcommand{\arraystretch}{1.2}
\begin{tabular}{||p{7cm}||p{7cm}|}
\hline
 \textbf{Question} & \textbf{Scale}  \\
\hline

\multicolumn{2}{|c|}{Video Annotation}\\
\hline
Do you think the person in the video is expressing anti-establishment views? & 4-point (from Y\textit{es, they are \textbf{definitely} expressing anti-establishment views} to
\textit{No, they are \textbf{definitely not} expressing anti-establishment views})
  \\

\hline
\multicolumn{2}{|c|}{Comment Annotation}\\
\hline
 Do you think the comment agrees with the information in the video? 
& 5-point (from \textit{Yes, I think this comment \textbf{definitely agrees} with the information in the video} to 
\textit{No, I think this comment \textbf{definitely does not agree} with the information in the video}  )
\\
\hline
\end{tabular}
\caption{Schematic of annotation task.}
\label{tab:dimensions_questions}
\end{table*}

\subsection{Machine annotation} 
After collecting labels from human annotators and aggregating the labels to arrive 
at a single label per example, 
we next employ machine learning to label the remaining instances in the dataset.

\subsubsection{Utilizing categorical information}
We also experiment with an additional feature representation for this task. 
The videos are collected to be representative of 
three distinct categories of content: conspiracy theories, 
finance, and wellness.  
As each category may have distinct linguistic signals which suggest AES 
we may benefit from representing  categorical information. 

In both the baseline and deep learning settings 
we incorporate categorical information directly. 
That is, in the baseline setting 
we concatenate a one-hot feature vector which encodes a video's category with the text-based feature vector. 
In the deep learning setting, 
we train an embedding layer which also takes a category vector as input.

\section{Empirical Evaluation}
We collect posts from the social media platforms TikTok and YouTube. 
First, we describe our data collection process. 
Next, we evaluate both the human and machine 
annotations. 

\section{Dataset}
\begin{table}[h!]
\resizebox{\columnwidth}{!}{%
    \centering
    \begin{tabular}{|c|ccc|c|}
        \hline
         & \conspiracy & \textsc{Finance} & \well & Total \\
        \hline    
        Number of videos & 8,439& 8,956 & 9,388 & 26,783\\     
        Number of comments &96,217 & 62,154& 47,978& 206,349\\  
        \hline
    \end{tabular}
    }
    \caption{Distribution of videos and comments across three categories in the datasets: \conspiracy, \finance, and \well. The primary TikTok dataset comprises a total of 26,783 videos and 206,349 comments.}
    \label{tab:dataset}
\end{table}
We collect $26,783$ video posts with $206,350$ comments using the TikTok Research API. 
Our goal was to collect posts oriented around the three themes
of finance, wellness, and conspiracy theories in recent years. 
Thus, we set the following parameters in the API requests: 

\begin{itemize}
    \item \textbf{Key words}: the post could contain any of the keywords from the sets of \textsc{Finance}, \textsc{Wellness}, \textsc{Conspiracy}.
    \item \textbf{Timeline:} the post was made between January 1, 2022, and December 31, 2023  for \textsc{Finance}, \textsc{Wellness}, \textsc{Conspiracy}.
    \item \textbf{Location:} the post was made in the United States. 
\end{itemize}

Where \textsc{Conspiracy} contained the phrases: \textit{``conspiracy'', ``flatearth'', ``propaganda'', ``illuminati''}, \textsc{Finance} contained the phrases: \textit{``finance'',``stocks'', ``crypto'', ``realestate''}, and
\textsc{Wellness} contained the phrases: \textit{``wellness'',``health'', ``selfcare'', ``fitness''}. 
To derive the phrases in each set we 
began with three umbrella terms: \textit{conspiracy}, \textit{finance}, and \textit{wellness}.  
Then we created a seed set of a few thousand posts which contained each 
umbrella term. We then chose common words within the seed set to add to 
one of the three sets of \textsc{Finance}, \textsc{Wellness}, or \textsc{Conspiracy}.
When there was a word which appeared to return posts unrelated to the target concept,
we did not add it to the set. 
Overall this process produced the dataset
shown in \tabref{tab:dataset}.

We describe the dataset at different stages of the analysis 
in \tabref{tab:funnel}. 
In the first phase (\textbf{Data Collection and Transcription}), we download videos from TikTok in the process described above.
This dataset consists of only  those videos which we were able to transcribe.
Next, in the \textbf{Data Filtering} phase, 
we retain only those videos such that length of the transcription and the video
description was at least 40 tokens, and the language of the text 
was detected as English\footnote{To detect the language of the text we used the spaCy package \citep{honnibal2020spacy}}. 
In the \textbf{Data Annotation} phase, 
we select a sample of video-comment pairs, where each comment to be annotated 
contains at least 10 tokens. 
Finally, we clean this dataset to only contain high quality annotations (\textbf{Annotated Data Filtering}).
To do so, we only considered annotations with at least three labels 
from the best annotators (see Human Annotation - Empirical Details). 
The \textbf{Gold Set} dataset is a subset of videos 
which the domain expert on the team annotated in terms of AES. 

\begin{table}[h!]
 \resizebox{\columnwidth}{!}{ 
\centering
\begin{tabular}{|c|c|c|}
\hline
\textbf{Stage} & \textbf{Videos} & \textbf{Comments} \\ \hline
Data Collection and Transcription  & 26,783 & 206,350 \\ \hline
Data Filtering & 14,261 & 129,996 \\ \hline
Data Annotation & 816 & 890 \\ \hline
Annotated Data Filtering  & 616 & N/A \\ \hline
Gold Set & 103 & N/A \\ \hline
\end{tabular}
}
\caption{For each state of data processing we display the number of videos and comments available. 
We did not consider the annotation of comments when filtering the annotated data, 
nor did we annotate comments in the Gold Set creation.}
\label{tab:funnel}
\end{table}

\subsection{Human Annotation - Empirical Details }
In the annotation task, annotators were asked to label 10 video-comment pairs.
Recall that, before starting the task, they were required to pass an assessment consisting of 4 questions, and they needed to achieve a perfect score to qualify for the task.
Moreover, to maintain high-quality annotations, we included two attention checks within the task. Annotators were required to pass both attention checks to successfully complete and submit their annotations. 
Ultimately, 39 annotators provided valid annotations, meeting all the requirements. These annotators were reimbursed at an average rate of \$21.90 per hour. Each video-comment pair was independently annotated by three different annotators.

To validate the human annotation we annotate a small gold set of 103 videos. We then inspect the precision, recall, F1 Score, and accuracy of the annotations under different aggregation schemes. 
In \tabref{tab:gold_set Macro-f1} we 
see that the best results are obtained 
with the Dawid Skene aggregation method. 
Additionally, we see that we obtain reasonable performance with a precision of 0.556. 
Thus, for the remainder of the paper we use the aggregate labels as determined by Dawid Skene. 

\begin{table*}[h!]
\centering
\small
\renewcommand{\arraystretch}{1.2}
\begin{tabular}{|c|c|c|c|}
\hline
\textbf{Aggregation Method} & \textbf{Dawid Skene \citep{dawid1979maximum}} & \textbf{MACE \citep{hovy2013learning}} & \textbf{Majority Vote} \\
\hline

\textbf{Precision} & 0.556 & 0.528 &\textbf{0.562}  \\

\hline
\textbf{Recall} & \textbf{0.909} & 0.864   & 0.818    \\
\hline
\textbf{F1 Score (binary)} & \textbf{0.690} & 0.665 & 0.667   \\
\hline
\textbf{F1 Score (macro)} & \textbf{0.784} & 0.760 & 0.774   \\
\hline
\end{tabular}
\caption{Performance comparison of different aggregation methods, evaluated using a gold standard set of size 103. Metrics include Precision, Recall, Binary F1 Score, and Macro F1 Score. In general, Dawid Skene does the best among these three methods and hence 
we used this aggregation method for the remainder of the paper. 
}
\label{tab:gold_set Macro-f1}
\end{table*}

\subsection{Machine Annotation - Implementation Details}
We treat our dataset as a text dataset. 
That is, for each video we produce a transcript using WhisperX \cite{bain23_interspeech}.
We then concatenate the video description to the video transcript 
to create a single text document for each video. 
Given this text dataset, 
we employ language models to annotate the remainder of the dataset. 

We would like to exploit the general knowledge stored in language models
which have been trained on large quantities of text.
Thus, we use BERT \citep{bert2019}, DistilBERT \citep{distilbert2020}, RoBERTa \citep{roberta2019} and DistilRoBERTa \citep{distilbert2020} as state-of-the-art
pretrained language models (PLMs). 
Additionally, 
we compare these approaches to two baselines: 
SentenceBERT \citep{sbert2019} plus SVM and SentenceBERT \citep{sbert2019} plus LightGBM \citep{lightgbm2017}. 

As the structure of AES content will likely vary with the category,
 we experiment with the inclusion of category specific information. 
 The way we incorporate such information depends on the model. 
For the SVM and LightGBM models, we one-hot encoded the category information and directly included it as a feature for training the models. 

For the PLMs, we pass texts to a PLM which outputs textual representations. To incorporate categorical information, we generate categorical embeddings.
That is, we train an Embedding layer which takes a category vector as input and outputs a dense embedding.
These category embeddings are concatenated with the PLM-derived representations.
The combined feature vector is then passed through a fully connected layer that maps the concatenated representations to a binary label space. The entire model, including the PLM, category embeddings, and the classification layer, is trained jointly in an end-to-end fashion to optimize performance on the binary classification task. This approach enables the model to effectively integrate semantic and categorical information for improved prediction accuracy.
The combined textual and categorical deep learning models are implemented in PyTorch \citep{paszke2019pytorchimperativestylehighperformance}.

For each case, we select hyperparameters by performing a stratified 5-fold cross-validation grid search on the training set, consisting of a total of 500 videos. Each grid search is conducted using three distinct random seeds, and the hyperparameters achieving the best performance across these seeds are selected.
The results on the held-out test set of 116 videos, averaged across the three random seeds, are reported in \tabref{tab:ResultTable}.

\begin{table*}[h!]
\centering
\small
\renewcommand{\arraystretch}{1.2}
\begin{tabular}{|c|c|c|c|c|c|c|}
\hline
\multicolumn{7}{|c|}{Pure Text}\\
\hline
\textbf{Metrics} & \textbf{SentenceBERT + SVM} & \textbf{LightGBM} & \textbf{BERT} & \textbf{DistilBERT} & \textbf{RoBERTa} & \textbf{DistilRoBERTa} \\
\hline
\textbf{Precision} &  $0.619_{0.000}$ & $\mathbf{0.833_{0.000}}$
 & $0.476_{0.006}$  & $0.501_{0.011}$ & $0.537_{0.051}$ &  $0.520_{0.010}$\\
\hline
\textbf{Recall} & $\mathbf{0.765_{0.000}}$ & $0.441_{0.000}$ & $0.580_{0.026}$  & $0.570_{0.009}$ & $0.755_{0.067}$ & $0.614_{0.023}$\\
\hline
\textbf{F1 Binary} & $\mathbf{0.684_{0.000}}$ & $0.577_{0.000}$ & $0.521_{0.007}$  & $0.533_{0.006}$ & $0.617_{0.019}$ & $0.562_{0.006}$\\
\hline
\textbf{Accuracy} & $0.793_{0.000}$ & $\mathbf{0.810}_{0.000}$ & $0.731_{0.004}$  & $0.747_{0.007}$ & $0.721_{0.046}$ & $0.758_{0.006}$\\
\hline
\multicolumn{7}{|c|}{Text + Category}\\
\hline
\textbf{Metrics} & \textbf{SentenceBERT + SVM} & \textbf{LightGBM} & \textbf{BERT} & \textbf{DistilBERT} & \textbf{RoBERTa} & \textbf{DistilRoBERTa} \\
\hline
\textbf{Precision} &  $0.553_{0.000}$ & $0.600_{0.000}$
 & $0.585_{0.011}$  & $0.562_{0.032}$ & $\mathbf{0.659_{0.009}}$ &  $0.555_{0.042}$\\
\hline
\textbf{Recall} & $0.765_{0.000}$ & $0.265_{0.000}$ & $0.676_{0.017}$  & $0.657_{0.020}$ & $\mathbf{0.814_{0.020}}$ & $0.725_{0.043}$\\
\hline
\textbf{F1 Binary} & $0.642_{0.000}$ & $0.367_{0.000}$ & $0.627_{0.013}$  & $0.604_{0.022}$ & $\mathbf{0.728_{0.013}}$ & $0.624_{0.012}$\\
\hline
\textbf{Accuracy} & $0.750_{0.000}$ & $0.733_{0.000}$ & $0.764_{0.008}$  & $0.747_{0.020}$ & $\mathbf{0.822_{0.008}}$ & $0.741_{0.028}$\\
\hline
\end{tabular}
\caption{Comparison of model performance across two experimental setups: one using pure text features and the other incorporating text with category embeddings. Metrics evaluated include average Precision, Recall, F1 Binary, and Accuracy, along with their respective standard errors, derived from experiments conducted with three distinct random seeds. In the pure text setup, SentenceBERT + SVM achieves the highest F1 Binary (0.684) and Recall (0.765), while LightGBM excels in Precision (0.883) and Accuracy (0.810). With category embeddings, RoBERTa demonstrates the strongest performance across all metrics.
}
\label{tab:ResultTable}
\end{table*}

The inclusion of category specific information was helpful for this task. 
We inspect the binary F1-score, as an informative measure in the setting with class imbalance. 
The best F1-score of $0.728$ is achieved by RoBERTa with the inclusion of category-specific information. 

\xhdr{Training Details:} For the RoBERTa model in the Text + Category setting, the final hyperparameters are as follows: PLM: roberta-large, learning rate: $1 \cdot 10^{-5}$,  epochs: $40$, size of the categorical embeddings: $32$, and class weights for positive and negative labels: $0.35$ and $0.65$, respectively. All models were trained on an AMD EPYC 7763 64-Core CPU and three NVIDIA RTX A5000 GPUs.

\section{Research Questions and Findings}
Next, we return to our original research questions. 
First, we annotate the entire dataset using the best 
machine learning model found using the validation set and 
trained on the human annotations: RoBERTa with categorical information.
Then, we use this entire dataset of both human and machine annotations to answer the research questions. 

\begin{figure}[h] 
    \centering 
    \includegraphics[width=0.39\textwidth]{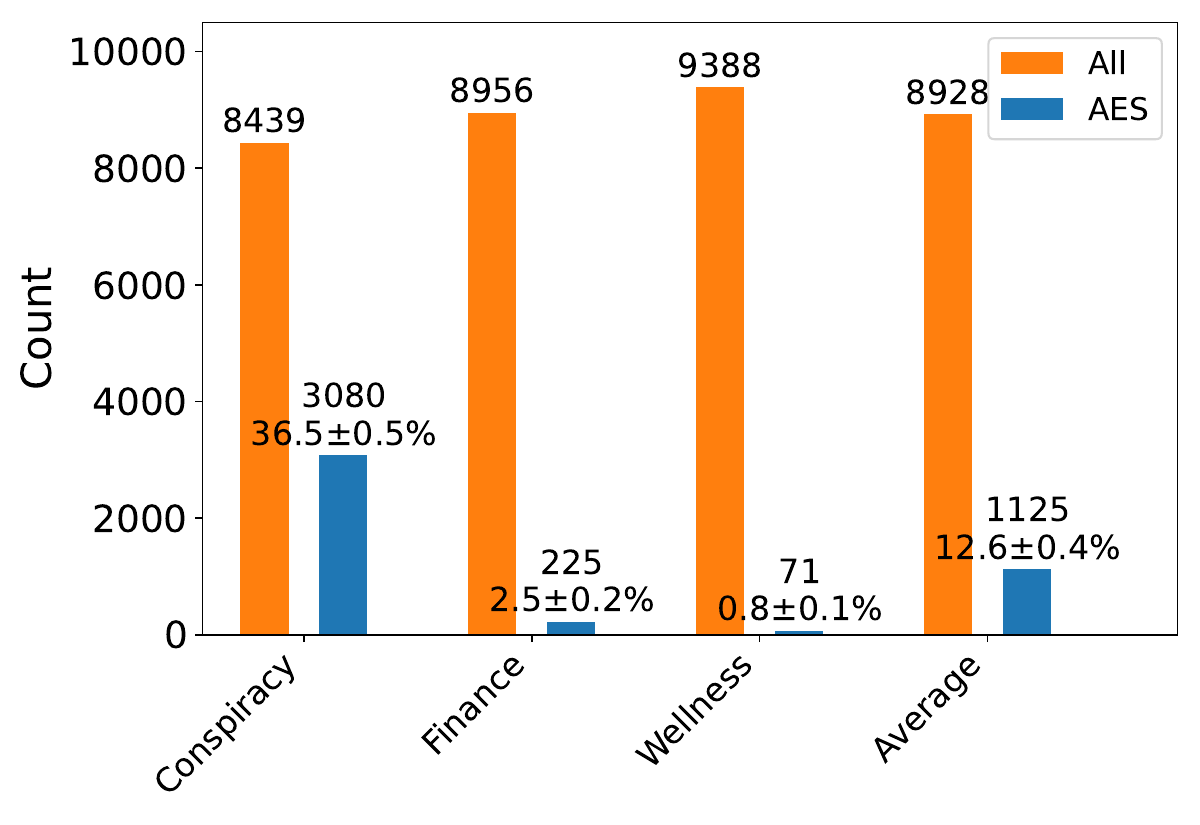} 
    \caption{
    Total video counts are shown above the bars. \conspiracy{} has the highest AES proportion (45.1\%), followed by \finance{} (4.3\%) and \well{} (1.3\%), with an overall average of 16.1\%. Confidence intervals across 3 seeds.
    }
    \label{fig:aesbycategory} 
\end{figure}

\xhdr{How does the use of AES differ by post topic on TikTok?}
We see in \figref{fig:aesbycategory} that the prevalence of AES sentiment varies widely across the categories. 
Results show the greatest proportion of AES is found in the \conspiracy{}  category, and we see little AES content in the categories of \finance{} (less than 5\%)  and \well{} ($\sim\!\!\!1\%$). 
The high proportion of AES content in \conspiracy{} serves to validate our approach. 

\begin{figure*}[h!]
    \centering 
    \includegraphics[width=.8\textwidth]{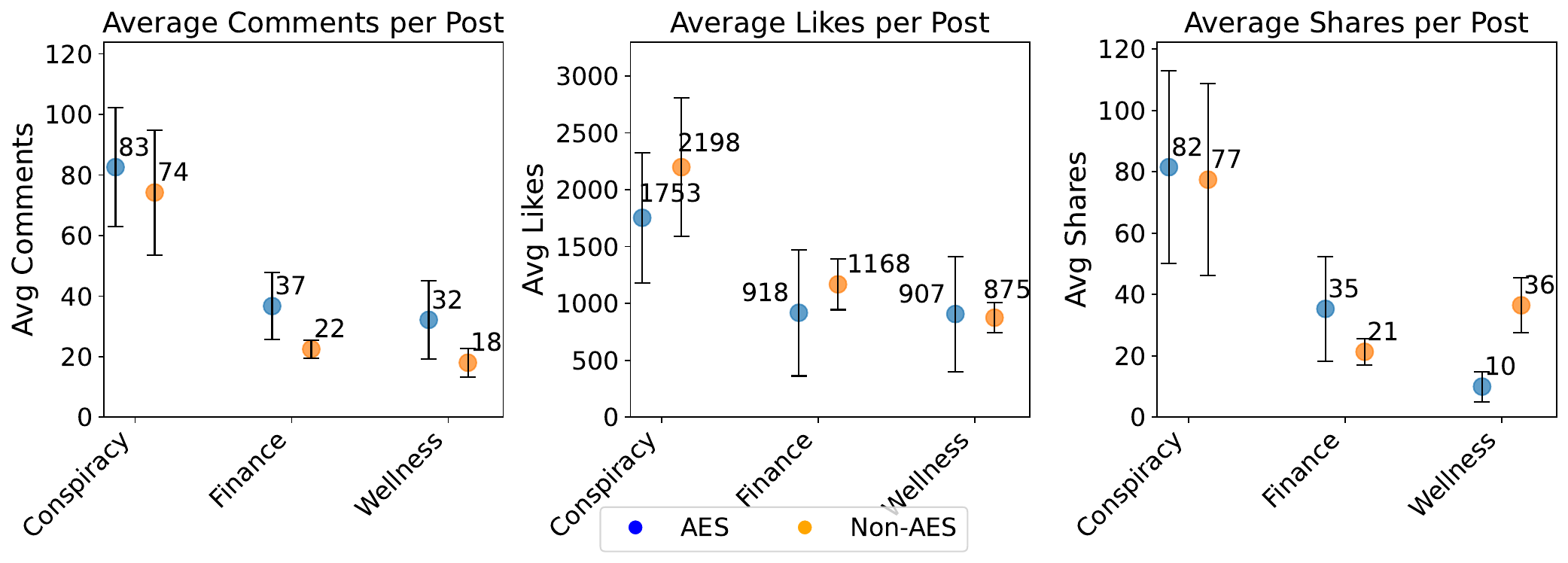} 
    \caption{
Engagement metrics (Comments, Likes, Shares) per post, categorized by AES (Anti-Establishment Sentiment) and Non-AES content across \conspiracy, \finance, and \well{} topics. AES posts show lower engagement in all metrics for \conspiracy{}; AES posts show higher engagement in Comments and Shares but lower engagement in Likes for \finance; and AES posts show higher engagement in Comments but lower engagement in Likes and Shares for \well.}
    \label{fig:engagement} 
\end{figure*}

\xhdr{How do users engage with content with AES and without it and how does this engagement vary by topic?}
In \figref{fig:engagement}, 
we see differences in some forms of engagement depending on whether a post contains AES or not. 
For example, in the area of \finance{} AES content receives slightly more comments and shares, 
than non-AES content. In the area of \well{} it receives more comments per post, 
but far fewer shares per post. 
In all three categories it receives fewer likes per post. 

In the annotation task, 
annotators are asked to determine if a comment made in response 
to a post agrees with the content of the post, disagrees with the content of the post, 
or seems to be irrelevant. 
We now use those labels to understand if viewers of posts with AES tend to agree or disagree 
with the content in the posts. 
Here, we take the majority vote for each comment label, and if there is a three-way tie, 
or if the majority vote assigns the irrelevant label, we mark the agreement of the comment 
as unclear. 

\begin{figure}[h] 
    \centering 
    \includegraphics[width=0.32\textwidth]{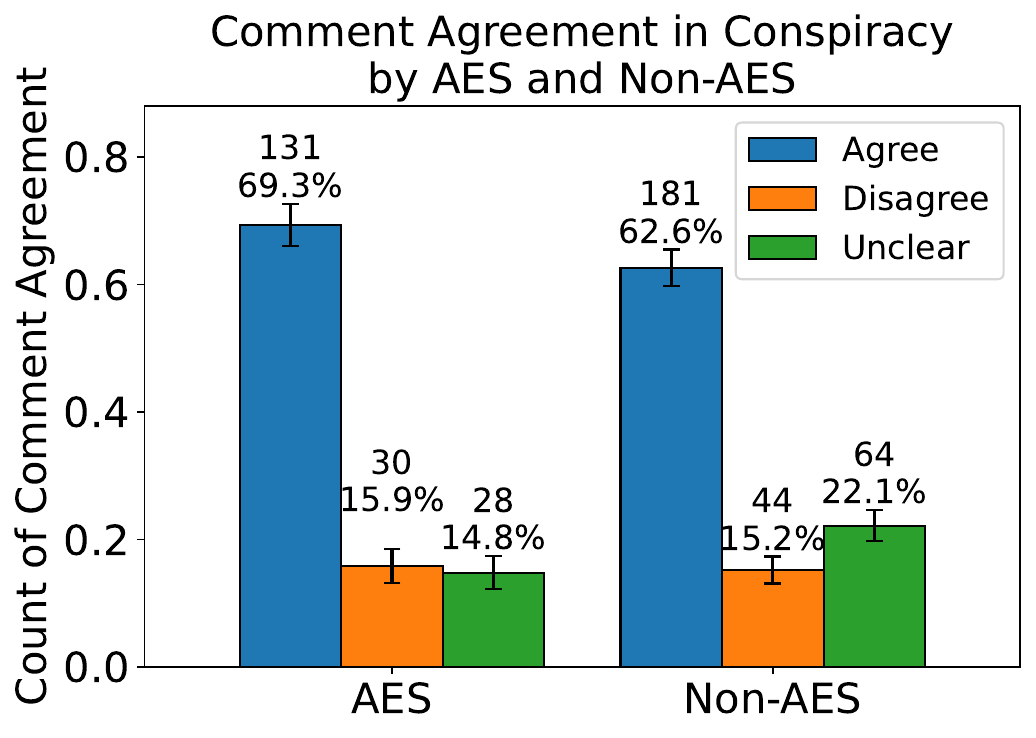} 
    \caption{Distribution of comment agreement in the \textsc{Conspiracy} category for AES (anti-establishment) and Non-AES labels. AES comments show a higher agreement rate (69.3\%) compared to Non-AES comments (62.6\%). This suggests that engagement with conspiracy content is not-driven by users refuting conspiracy theories.}
    \label{fig:conspiracy_comments_agreement} 
\end{figure}

In \figref{fig:conspiracy_comments_agreement}, 
we analyze a subset of comments made in response to TikTok posts
in the \conspiracy{} category (the only category with sufficient comment annotations to warrant analysis). 
Here, we see that in aggregate commenters are agreeing with AES content as much, 
if not more so, than non-AES content. 

\xhdr{Which institutions are most likely to be the target of AES?}
A particularly important question towards understanding anti-establishment sentiment 
is which institutions are the target of such sentiment.
To address this question we hand coded a sample of 50 posts
which were predicted to contain AES. 

In particular we coded each post as to whether it contained the following institutions: 

\begin{itemize}
    \item \textbf{US Government:} A mention that the US government is not to be trusted. 
    \item \textbf{Politicians:} A mention that politicians are not to be trusted/are corrupt. 
    \item \textbf{US Healthcare:} A mention that there is something fundamentally wrong/untrustworthy  about the US Healthcare system.
    \item \textbf{Big Pharma:} A mention that there is something fundamentally wrong/untrustworthy about pharmaceutical companies. 
    \item \textbf{Big Banks:} A mention that there is something fundamentally wrong/untrustworthy about financial institutions.
    \item \textbf{NASA:} A mention that NASA is corrupt and/or lies to the American public. 
    
\end{itemize}

\begin{figure}
    \centering
    \includegraphics[width=0.99\linewidth]{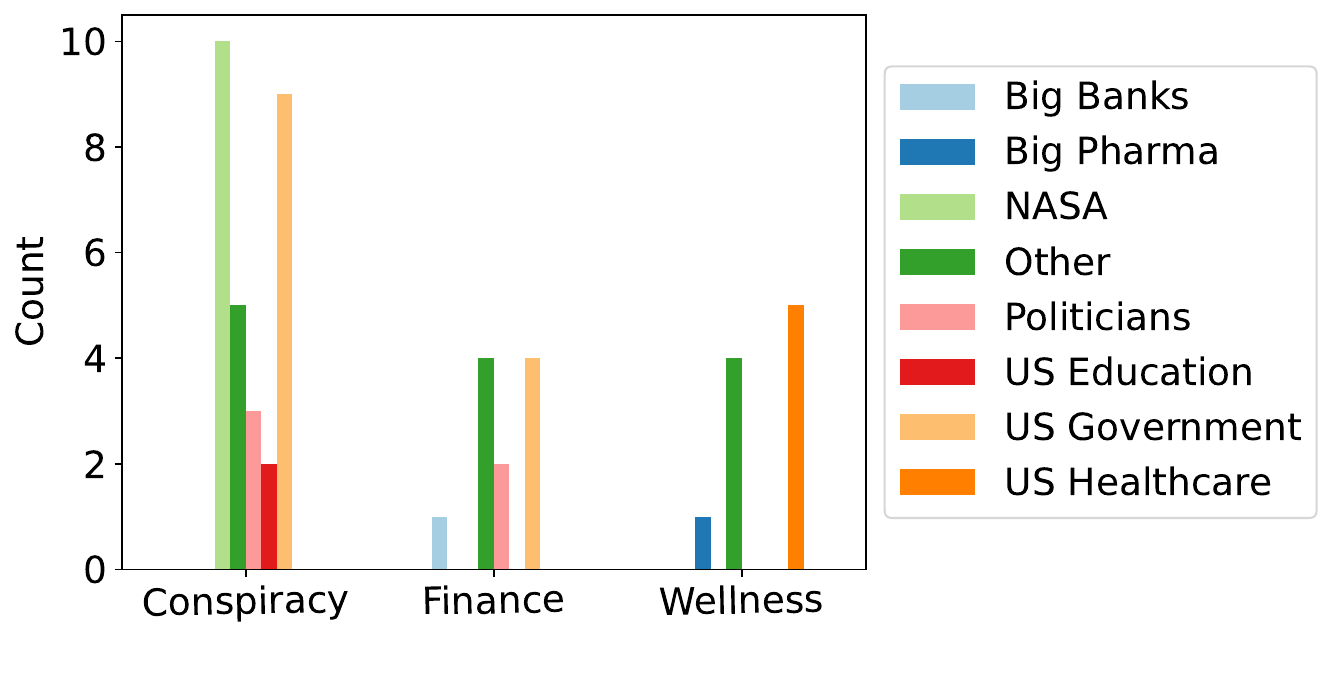}
    \caption{We see that the institutions which are the targets of AES vary by category. 
    The most common target across categories is the US Government, while the Conspiracy videos with many Flat Earth videos target NASA, the Finance videos have diffuse targets, and the Wellness videos target the US Healthcare system.
    }
    \label{fig:institutions}
\end{figure}

In \figref{fig:institutions}, 
we offer a cursory glimpse into which institutions are discussed in these posts across the different categories. We manually code 50 posts, thus we don't present this analysis as representative, but 
as a way of understanding  how the categories likely differ. 
For example, US Healthcare is mentioned only in Wellness posts, but posts referring to the US Government appear both in conspiracy and finance posts. When a particular US Government body was mentioned (like NASA) we noted this, and results suggest  that references to the government in the abstract are more common than accusations targeted at specific bodies. 

\xhdr{Is the relationship between post topic and AES content similar across social media platforms?}
To address this question we use the same keywords for each of the three topics and collect $\sim$80 posts using the YouTube API for each set of keywords for a total of 1013 posts.
We set the region code to the United States, the language to English, and the timeline to be between January 1, 2023 and December 31, 2024.
This sample allows us to see how AES content differs from TikTok to YouTube for the same set of topics.

We find that the ranking of the three topics in terms of AES prevalence is the same. That is, 
20\% of the conspiracy posts, 4\% of the finance posts, and less than 1\% (.06\%) of the wellness posts contain AES.
This shows that users interested exclusively in wellness would be least likely to encounter AES on either platform. Moreover, like on TikTok, users primarily interested in finance would be slightly more likely to encounter AES than users interested in wellness.  
Our sample also suggests that users who exclusively use YouTube would be less likely to encounter AES in conspiratorial content than those
who exclusively use TikTok.
This  may be a result of YouTube's policies to suppress problematic content, 
 \citep{buntain2021youtube}. 

\subsubsection*{For You Page Data}
We additionally collect data by mimicking
human users who would log into the FYP  on TikTok. To do so we create 48 sock puppet accounts. These accounts were created in the months of September and October 2024 and browsed content from October 2024 to December 2024 (35 days on average). To mimic a user who may be interested in a few topics, and the news, when each account is created it is randomly assigned  6 lifestyle accounts and 6 news channel accounts to follow (shown in the Appendix in \tabref{tab:news} and \tabref{tab:influencers}).  Each day each account logged on to their FYP at a randomly assigned time between 7:00 am and 9:00 am EST. They then logged on again exactly twelve hours later. In each session an account paused scrolling to watch each video with a 50\% probability. 

Unlike the other datasets
this dataset is not filtered to particular keywords. It thus offers a more generalized view into how often AES content is shown to a user who only visits the FYP, 
rather than seeking topic-driven content.

All together, this procedure produced 37,012 videos. The research team annotated 349 of these videos for AES. We then trained a RoBERTa model as that performed the best in the other experiments. With RoBERTa, we achieved a binary F1 of 0.822, a precision of 0.889, a recall of 0.833, and a accuracy of 0.990. 
We thus proceeded to infer the labels of the remaining videos utilizing this trained model.

We found that less than 1\% (0.48\%) of all posts expressed AES. This again shows that the extent to which a user will be exposed to AES on TikTok 
depends on their interests. While a  user who periodically browses the FYP may see very little AES content, a user interested in conspiracy theories will see much more.

\input{lingustic_rq}

\input{puppet_data}

\section{Discussion}
This study systematically examines the prevalence of anti-establishment sentiment on TikTok across a range of topics that are not explicitly political. Prior research has shown that social media platforms tend to reward and amplify political content that promotes anti-establishment sentiment \citep{milmo2021twitter, starbird2023influence, tripodi2024your, zadrozny2021carol}, and other work has suggested that even apolitical content on social media may facilitate and encourage anti-establishment views \citep{chinn2026does, hasell2023political} as influencers and content creators position themselves as voices of expertise and authority in social media \citep{wellman2024friend}. Our results suggest that AES in posts about finance and wellness is relatively uncommon. We found AES in an estimated 4\% of all posts we collected in the domain of \finance{}
and 1\% of all posts we collected in \well{}. 
In contrast, we find that 45\% of all \conspiracy{}
posts contained AES. Overall, this suggests that anti-establishment sentiment is not widespread in apolitical content on TikTok. 

However, such posts may generate more comments and shares than other types of content. For example, in \finance{} and \well{}, our results show that AES content is more likely to be commented on, and in \finance{} such content is more likely to be shared. Though we also note that across all three categories, including \conspiracy{}, AES content is liked less often. In regard to the comments, in \figref{fig:conspiracy_comments_agreement}, we find generally that comments agree with the posted content. Upon manual inspection we do see occasional posts which will ridicule for example, flat-earth theories, 
but in aggregate, in our sample of labeled data we see that comments tend to agree with the original post. 

Future work should consider both how anti-establishment sentiment might appear in a wider range of topics and the potential long-term effects of encountering such content. Though only a small portion of posts contained AES, there remain questions as to how influential AES content might be when users are repeatedly exposed over time.

We propose including a categorical representation with the textual representation of each post and find this approach to be useful. 
When such information was included, we see that RoBERTa achieves the best performance overall. 
This is unsurprising, as this is the one of the larger models we experiment with. 
Future work could focus on a single category, e.g. wellness, and train models specifically for that domain. 

Finally, we would like to note that the concept of AES is subtle. 
Annotating the Gold Set required much deliberation across the research team to ensure 
that our annotations were consistent and in line with what crowd workers could do. 
We see a line of future work seeking to understand how this term  can best be measured 
as it evolves.

\subsection{Limitations}
One significant limitation of this work is that we did not investigate the number of views for each post. 
However, this was not an analytical oversight, but an empirical decision. 
We found that the view data returned by the TikTok API 
was unreliable. 
That is, there were posts with 0 views but more than 0 likes. 
We researched this discrepancy and found a range of non-conclusive 
hypotheses. For example, it is possible that TikTok fails to reliably log 
these characteristics. 
It is also possible that TikTok manually sets the views of certain types of content or accounts to 0. 
However, we found no difference in the prevalence of this discrepancy across 
the three categories. 

\xhdr{Ethical Implications}
There may be concerns that  collecting posts made by individuals is a violation of privacy. However, the posts we accessed were all publicly available and we do not share any individual's data or content. For example, we do not publish data related to whether any individual post or content creator was labeled as anti-establishment.  Additionally, all annotators were provided with informed consent and an institutional IRB reviewed the study and ruled it to be exempt.

\section{Conclusion}
We find that AES content does occur across three types of content 
on TikTok,  conspiracy theories, finance, 
and wellness, to varying degrees. Such content is common in conspiracy content, 
but rare in the domains of finance and wellness. 
This result generalized to a sample of YouTube content as well. That is, we found that the relative prevalence of AES in these topics was the same on YouTube, while YouTube conspiracy content expressed less AES than conspiracy content on TikTok.
We also considered a collection of posts from TikTok's FYP. Here, we found that AES is expressed in less than 1\% of all content.
Thus, users will encounter different levels of exposure to AES, depending on their interests and behaviors. A user who browses the FYP with limited interaction will likely encounter very little AES content, and a user who seeks conspiratorial content will likely encounter it regularly.
However, better understanding the implications of such content requires further testing of the effects of exposure to AES in social media and examination over longer periods of time. With anti-establishment views increasing, and the increased prominence of political figures who hold anti-expert and anti-establishment views \citep{zadrozny2025},
it may be that such content will become increasingly prevalent. 

\bibliography{aaai2026}

\clearpage

\input{ethics}

\clearpage

\appendix
\section{Appendix}
\section{Visual Cues}
TikTok videos contain rich visual information above and beyond what can be captured through transcribed audio. 
Here, we analyze whether there are differences in how content creators shoot their videos by topic. To do so we tag 738 videos according to their visual style. 
For example, as authenticity is an important component of AES, we might expect the conspiracy videos (the topic with the highest proportion of AES) to more often contain a person speaking directly to the camera. 
Alternatively, we may expect the opposite, as building authority can also be a component of AES. 

In \tabref{tab:vis}, 
we see that people are least likely to speak directly to the camera in the conspiracy content. This perhaps reflects that fact that the content creators in finance and wellness are more likely to be building a personal brand where centering the narrative around themselves is important. 
We also see that these categories are less likely to use embedded media than the conspiracy topic. 

Next, we inspect how often the posts feature a person speaking directly to the camera in the conspiracy content that expresses AES and does not. 
Here, we see that in the conspiracy videos that express AES a person speaks directly to the camera 39\% of the time and in the videos that do not express AES a person speaks directly to the camera 32\% of the time. Thus, we see that overall conspiracy videos are less likely to feature a single person speaking to the camera than the other topics, however, AES content is slightly more likely to feature this point of view. This slightly supports the hypothesis that AES content creators choose visual cues that build authenticity, however, authenticity may be a less important factor for AES content than for content where personal brand building is important. 

\begin{table}[h!]
 \resizebox{\columnwidth}{!}{ 
    \begin{tabular}{|l|r|r|r|}
    \hline
         & Conspiracy & Finance & Wellness \\
         
         Speaking directly to camera& .35 &.57 &.46 \\
         Speaking but not to camera &.34 &.26 &.31 \\
         Embedded media &.10 &.05 &.005 \\
         Text overlay with music & .06& .04& .07 \\
         Other & .15&.08 & .145\\
         \hline
    \end{tabular}
    }
    \caption{We inspect the ways the videos are shot for each of the three topics of conspiracy, finance, and wellness. Each entry is the proportion of content within this category that employs this visual style. \label{tab:vis}}
\end{table}

\section{Information on Accounts Followed by Sock Puppets (For You Page Data)}
We create 48 sock puppet accounts which interact with the For You Page. 
When the accounts are created they follow both news channels and lifestyle channels. 
The list of news channels is shown in \tabref{tab:news} and the list of lifestyle channels is shown in \tabref{tab:influencers}.

\begin{table}[htb!]
    \centering
    \input{news}
    \caption{Each account followed 6 news channels from the list above, where the channels were selected at random. Reliability scores and bias scores are extracted from the Media Bias Chart \citep{adfontesmedia}. Reliability scores for articles and shows are on a scale of 0-64. Scores above 40 are generally good. Bias scores for articles and shows are on a scale of -42 to +42, with higher negative scores being more left, higher positive scores being more right, and scores closer to zero being minimally biased, equally balanced, or exhibiting a centrist bias. Typically, a publication would be considered centrist if the score is between -10 and +10, left-oriented if the score is -10 or less and right-leaning if the score is +10 or more. Thus, these accounts are all considered to be centrist.}
    \label{tab:news}
\end{table}

\begin{table}[htb!]
    \centering
    \input{influencers}
    \caption{Lifestyle influencers followed by the puppet accounts. Each account was assigned 2 topics and three followers from each topic.}
    \label{tab:influencers}
\end{table}

\end{document}

%% file: lingustic_rq.tex
\subsubsection{How do videos with AES content use linguistic style to
create divisions between the People and elites?} 

Returning to our definition of AES as an expression of a perceived power struggle between the People and elites (or the establishment) \citep{uscinski2021american},
we would like to see how this tension surfaces through the linguistic style of content creators on social media. 
We propose that such divisions will occur through three primary themes: \begin{itemize}
    \item Authenticity: Creators of AES content may portray the people as having real authenticity and elites as lacking connections to the lives and circumstances of ordinary individuals' lived experiences. 
    \item Authority: Creators of AES content may use a style which vests themselves with authority and the elites as lacking the real authority of the people. 
        \item Morality: Creators of AES content may use a moralistic style, speaking to the morality of the people and the amorality of elites. Religious themes will be evoked with the people being on the side of  good faith and the elites being portrayed as lacking faith. 
\end{itemize}

To explore these three themes we employ the most recent version of the Linguistic Inquiry and Word Count (LIWC2) software \citep{boyd2022development}. LIWC analyzes text by calculating the proportion of words that fall into predefined linguistic categories. For each of the three themes we inspect specific cues which signal the use of the theme. 
\tabref{tab:linguistics}  shows these themes and the factors and variables assigned to them.

Authenticity reflects the extent to which language is personal and self-revealing, as opposed to detached and guarded \cite{boyd2022development}. Language with high authenticity scores is characterized by a greater use of I-words, present-tense verbs, and relativity words (e.g., \textit{old}, \textit{far}, \textit{here}), and a reduced use of third-person pronouns (e.g., \textit{she}, \textit{he}) and discrepancy words. 

\begin{table}[h!]
 \resizebox{\columnwidth}{!}{ 
\centering
\begin{tabular}{|c|c|c|c|c|c|}
\hline
\textbf{Theme} &\textbf{Factors} & \textbf{Variables} & \textbf{Labels} & \textbf{Mean} & \textbf{SE} \\ \hline

Authenticity & Persuasive Tone & Authenticity & Non-AES & 48.7 & 0.25 \\ 
& &  & AES & 43.8 & 0.47  \\ 
&In-out-group & First-person singular & Non-AES & 3.91 & 0.03 \\ 
& &  & AES & 2.34 & 0.05 \\ 
 & & First-person plural & Non-AES & 0.81& 0.01 \\ 
& &  & AES & 1.39 & 0.03 \\ 
&   & Third-person plural & Non-AES & 0.49& 0.01 \\ 
& &  & AES & 1.39 & 0.03  \\ \hline
Authority &Relevance & Clout & Non-AES & 56.3 & 0.26 \\ 
& &  & AES & 62.1 & 0.45  \\ 
& & Power-related & Non-AES & 0.81 & 0.01 \\ 
& &  & AES & 1.98 & 0.04  \\ \hline
& Gender & Male reference & Non-AES & 0.62	& 0.01 \\ 
& &  & AES & 0.82 & 0.02 \\ 
& & Female reference & Non-AES & 0.53	& 0.01 \\ 
& &  & AES & 0.29 & 0.02 \\ \hline
Morality& Relevance  & Religion-related & Non-AES & 0.31 & 0.01 \\ 
& &  & AES & 0.79 & 0.04  \\ 
& & Death-related & Non-AES & 0.09 & 4e-3 \\ 
& &  & AES & 0.18 & 9e-3  \\ \hline
\end{tabular}
}
\caption{Summary of linguistic cue measurements across different variables and labels. The table categorizes linguistic features into four main factors: Persuasive Tone, Relevance, In-Group vs. Out-Group, and Gender. For each variable within these factors, the mean and standard error (SE) values are reported separately for Non-AES and AES labels.}
\label{tab:linguistics}
\end{table}
Authority reflects the extent to which language asserts epistemic knowledge through certainty and dominance. We use the LIWC variable Clout  to measure authority in the language of a post. A higher Clout score indicates a more powerful and confident language style, which is characterized by a greater use of we words and social words, alongside a reduced use of I words, negations (e.g., \textit{no}, \textit{not}), and swear words. 
Power-related words, such as \textit{own}, \textit{order}, \textit{allow}, and \textit{power}, reflect themes of dominance, control, and influence.

Morality reflects the extent to which language is used to position events or ideas as in the interest of society or the good of the people, as opposed to harm or evil. 
We use the LIWC variable Religion-related words, such as \textit{god}, \textit{hell}, and \textit{church}, indicate discussions of spirituality, beliefs, or religious practices, often reflecting cultural or moral perspectives. Death-related words, including \textit{die}, \textit{kill}, and \textit{coffin}, highlight themes of mortality, loss, and existential concerns, which can signal emotional intensity or warnings.

The linguistic analysis reveals distinct patterns between AES and non-AES content across key dimensions.
When it comes to Authenticity, we see that non-AES content demonstrates higher Authenticity (48.7 vs. 43.8), suggesting greater personalization through first-person narratives and present-tense language, a strategy linked to trust-building in lifestyle content \cite{hasell2023political}.
 In-out-group dynamics further highlight differences: AES posts favor collective identity (first-person plural: 1.39 vs. 0.81; third-person plural: 1.39 vs. 0.49) over individual perspective (first-person singular: 2.34 vs. 3.91), reinforcing the ``people vs. elites'' dichotomy central to anti-establishment rhetoric \cite{mudde2004populist}. This aligns with Klein's findings on social media posts promoting conspiracy theories, which similarly employ in-group and out-group language to create division, leverage emotional appeals to provoke and engage audiences, and reference religious doctrines as a means of validation \cite{klein2023loaded}. By strategically using these rhetorical devices, AES posts not only reinforce the ``people vs. elites'' dichotomy but also position themselves as credible and authoritative voices, effectively garnering trust and influence among their audiences.

When it comes to Authority, the Persuasive Tone metrics show AES posts exhibit significantly higher Clout scores (AES: 62.1 vs. Non-AES: 56.3), indicating a more authoritative and collective language style. This aligns with findings that anti-establishment creators position themselves as alternative authorities by adopting confident rhetoric to challenge institutional expertise 
 \cite{chinn2026does}.

When it comes to Morality, we see that religious-related terms are more prevalent in AES content (religion: 0.79 vs. 0.31), reflecting that some people tend to find reference from religious documents. (e.g., ``The earth is flat. You can find it in Bible.''). Death-related terms are also more prevalent in AES content (death: 0.18 vs. 0.09), echoing that AES spreads fear and anger by association with death. (e.g., ``Take a vaccine that could maybe make me die'').

Gender-related language shows minor disparities in ways which may relate to authority and morality: AES posts use more male references (0.82 vs. 0.62) and fewer female references (0.29 vs. 0.53), potentially reflecting gendered stereotypes in conspiratorial narratives (e.g., framing male figures as dominant antagonists). Research indicates that conspiracy theories often employ gendered language, with a tendency to reference male figures more frequently than female ones. Work analyzing gender representations in conspiracy discourse found that such narratives often reinforce connections between religiosity and masculinity, while relying on biological gender essentialism to define femininity \citep{fleckenstein2025representations}. The results from our study suggest that male figures are more prominently featured, potentially as dominant antagonists in AES posts.

In summary, we see that AES content creators use linguistic style to build their own authority and can draw on themes related to morality.  AES content creators score lower on the authenticity variable than non-AES content creators. While we hypothesize that these content creators communicate authenticity in different ways, this finding also creates opportunities for future study.

%% file: puppet_data.tex
\commentout{\subsection{AES Prevalence on For-You-Page}

\textcolor{green}{To explore the prevalence of, we experiment with the scraped data from For-You-Page. We deployed automatic bot to extract videos from the For-You-Page. ..... Since the Roberta was good in the previous experiments, we finetuned a Roberta-large model on the For-You-Page dataset.}

\begin{table*}[h!]
\centering
\small
\renewcommand{\arraystretch}{1.2}
\begin{tabular}{|c|c|}
\hline
\multicolumn{2}{|c|}{Pure Text}\\
\hline
\textbf{Metrics} & \textbf{RoBERTa} \\
\hline
\textbf{Precision} &  $0.889_{0.111}$\\
\hline
\textbf{Recall} & $\mathbf{0.833_{0.167}}$ \\
\hline
\textbf{F1 Binary} & $\mathbf{0.822_{0.097}}$ \\
\hline
\textbf{Accuracy} & $0.990_{0.005}$ \\
\hline
\end{tabular}
\caption{The results on the puppet dataset}
\label{tab:puppet_results}
\end{table*}}

%% file: ethics.tex
\section{Paper Checklist}

\begin{enumerate}

\item For most authors...
\begin{enumerate}
    \item  Would answering this research question advance science without violating social contracts, such as violating privacy norms, perpetuating unfair profiling, exacerbating the socio-economic divide, or implying disrespect to societies or cultures?
    \answerYes{}
  \item Do your main claims in the abstract and introduction accurately reflect the paper's contributions and scope?
    \answerYes{}
   \item Do you clarify how the proposed methodological approach is appropriate for the claims made?
    \answerYes{}
   \item Do you clarify what are possible artifacts in the data used, given population-specific distributions?
    \answerYes{See Dataset, Human Annotation and Machine Annotation}
  \item Did you describe the limitations of your work?
    \answerYes{See Limitations}
  \item Did you discuss any potential negative societal impacts of your work?
    \answerYes{we discussed the potential negative impacts of our work. We concluded that this work has no negative impacts and instead has positive impacts by adding new scientific analyses of an important societal problem.}
      \item Did you discuss any potential misuse of your work?
    \answerYes{}
    \item Did you describe steps taken to prevent or mitigate potential negative outcomes of the research, such as data and model documentation, data anonymization, responsible release, access control, and the reproducibility of findings?
    \answerYes{We were worried that human annotators may be exposed to troubling content. We reviewed the annotation with IRB, collected informed consent, and put a trigger warning before each post. Annotators always had the option to not annotate a post if the content was upsetting to them. }
  \item Have you read the ethics review guidelines and ensured that your paper conforms to them?
    \answerYes{}
\end{enumerate}

\item Additionally, if your study involves hypotheses testing...
\begin{enumerate}
  \item Did you clearly state the assumptions underlying all theoretical results?
    \answerNA{}
  \item Have you provided justifications for all theoretical results?
    \answerNA{}
  \item Did you discuss competing hypotheses or theories that might challenge or complement your theoretical results?
    \answerNA{}
  \item Have you considered alternative mechanisms or explanations that might account for the same outcomes observed in your study?
    \answerNA{}
  \item Did you address potential biases or limitations in your theoretical framework?
    \answerNA{}
  \item Have you related your theoretical results to the existing literature in social science?
    \answerNA{}
  \item Did you discuss the implications of your theoretical results for policy, practice, or further research in the social science domain?
    \answerNA{}
\end{enumerate}

\item Additionally, if you are including theoretical proofs...
\begin{enumerate}
  \item Did you state the full set of assumptions of all theoretical results?
    \answerNA{}
        \item Did you include complete proofs of all theoretical results?
    \answerNA{}
\end{enumerate}

\item Additionally, if you ran machine learning experiments...
\begin{enumerate}
  \item Did you include the code, data, and instructions needed to reproduce the main experimental results (either in the supplemental material or as a URL)?
   \answerYes{
Partially yes. Please refer to the Empirical Evaluation section. The code is open-sourced.}
  \item Did you specify all the training details (e.g., data splits, hyperparameters, how they were chosen)?
    \answerYes{See Machine Annotation - Implementation Details}
     \item  Did you report error bars (e.g., with respect to the random seed after running experiments multiple times)?
    \answerYes{See \tabref{tab:ResultTable}}
        \item Did you include the total amount of compute and the type of resources used (e.g., type of GPUs, internal cluster, or cloud provider)?
    \answerYes{}
     \item Do you justify how the proposed evaluation is sufficient and appropriate to the claims made?
    \answerYes{}
     \item Do you discuss what is ``the cost`` of misclassification and fault (in)tolerance?
   \answerNo{We see limited to no cost and went to great lengths to get the highest quality annotations we could.}

\end{enumerate}

\item Additionally, if you are using existing assets (e.g., code, data, models) or curating/releasing new assets, \textbf{without compromising anonymity}...
\begin{enumerate}
  \item If your work uses existing assets, did you cite the creators?
    \answerYes{}
  \item Did you mention the license of the assets?
    \answerNA{}
  \item Did you include any new assets in the supplemental material or as a URL?
    \answerNA{}
  \item Did you discuss whether and how consent was obtained from people whose data you're using/curating?
    \answerYes{See Human Annotation}
  \item Did you discuss whether the data you are using/curating contains personally identifiable information or offensive content?
    \answerYes{See Human Annotation and Findings}
\item If you are curating or releasing new datasets, did you discuss how you intend to make your datasets FAIR?
\answerNA{}
\item If you are curating or releasing new datasets, did you create a Datasheet for the Dataset?
\answerNA{}
\end{enumerate}

\item Additionally, if you used crowdsourcing or conducted research with human subjects, \textbf{without compromising anonymity}...
\begin{enumerate}
  \item Did you include the full text of instructions given to participants and screenshots?
    \answerYes{See Table \ref{tab:dimensions_questions}}
  \item Did you describe any potential participant risks, with mentions of Institutional Review Board (IRB) approvals?
    \answerYes{See Human Annotation}
  \item Did you include the estimated hourly wage paid to participants and the total amount spent on participant compensation?
    \answerYes{See Human Annotation}
   \item Did you discuss how data is stored, shared, and deidentified?
   \answerYes{}
\end{enumerate}

\end{enumerate}

%% file: news.tex
\begin{tabular}{llrr}
\toprule
author\_name            & author\_id         & Reliability & Bias  \\
\midrule
New York Times  & nytimes        & 41.04       & -8.07 \\
NBC             & nbcnews        & 42.80        & -5.64 \\
Washington Post & washingtonpost & 38.83       & -6.93 \\
PBS News        & pbsnews        & 43.32       & -4.05 \\
ABC             & abcnews        & 44.80        & -3.00    \\
CBS             & cbsnews        & 42.03       & -2.72 \\
NPR             & npr            & 43.09       & -4.17 \\
BBC News        & bbcnews        & 44.73       & -1.35 \\
Yahoo News      & yahoonews      & 40.94       & -5.63 \\
USA Today       & usatoday       & 40.86       & -4.06 \\
\bottomrule
\end{tabular}

%% file: influencers.tex
\begin{adjustbox}{width=0.55\textwidth,center}{
\def\sym#1{\ifmmode^{#1}\else\(^{#1}\)\fi}
\begin{tabular}{lll}
\toprule
Category                                   & author\_name                    & author\_id               \\
\midrule
Sustainability           & HomesteadDonegal        & mirendarosenberg     \\
                                           & Ken Russell             & kenforflorida        \\
                                           & Alaina Wood             & thegarbagequeen      \\
                                           & Steeze365Daily          & steeze365daily       \\
                                           & thesorrygirls           & thesorrygirls        \\
                                           & ReLauren                & relauren             \\
                                           & Reallaurinda            & reallaurinda         \\
                                           & Ashley Diedenhofen      & sciencebyashley      \\
                                           & TheNotoriousKIA         & thenotoriouskia      \\
                                           & Phil Sustainability     & philsustainability   \\
                                           & Brennan Kai             & brennan.kai          \\ \midrule
Wellness              &    Jessamyn $|$ The Underbelly Yoga                     & mynameisjessamyn     \\
                                           & Mari Llewellyn                        & marillewellyn        \\
                                           &  Arielle Lorre                       & ariellelorre         \\
                                           &          Dr. Will Cole               & drwillcole           \\
                                           &     Micheline Maalouf Therapist                    & micheline.maalouf    \\
                                           &  Staci Tanouye, MD                       & dr.staci.t           \\
                                           & Andrew Huberman                        & hubermanlab          \\
                                           &  Fiona                       & feelgoodwith\_fi     \\
                                           &   Steph Grasso, MS, RD                      & stephgrassodietitian \\
                                           &    Daniel                     & mrduku               \\ \midrule
DIY / Home improvement   & Molly Miller            & therenegadehome      \\
                                           & Bong Bain               & wildheartshome       \\
                                           & Lilly                   & thefurnituredoctor   \\
                                           & Christine Higg          & \_forthehome         \\
                                           & Joanna Gaines           & joannagaines         \\
                                           & Lone Fox                & lonefoxhome          \\
                                           & Kylie Katich            & kyliekatich          \\
                                           &  CASSMAKESHOME $|$ HOME \& DIY                       & cassmakeshome        \\
                                           &  Renovating Our Home                       & renovatingourhome    \\
                                           &  Jay Munee DIY                       & jaymuneediy          \\
                                           &   kelsey                      & kelseydarragh        \\
                                           &  Abby                       & abby\_roadhome       \\
                                           &  Contractor Ken                       & contractorken        \\
                                           &  ReallyVeryCrunchy                       & reallyverycrunchy    \\
                                           &  THE FLIPPED PIECE                       & theflippedpiece      \\
                                           &  Jeff Thorman                       & homerenovisiondiy    \\
                                           &  Bro Builds                       & bro\_builds          \\ \midrule
Tech                   & koharotv                & koharotvreal         \\
                                           & Tyler Morgan            & hitomidocameraroll   \\
                                           & Jimena con jota         & soyjimenaconjota     \\
                                           & CHIP                    & chip\_de             \\
                                           & Mark's Tech             & markstech            \\
                                           & TheAsianJC              & theasianjc           \\
                                           & Lucas VRTech            & lucas\_vrtech        \\
                                           & Marques Brownlee        & mkbhd                \\
                                           & Unbox Therapy           & unboxtherapyofficial \\
                                           & Austin Evans            & austintechtips       \\
                                           & iJustine                & ijustine             \\
                                           & Kevin Stratvert         & kevinstratvert       \\
                                           & Sara Dietschy           & saradietschy         \\ \midrule
College Sports         & Olivia Dunne            & livvy                \\
                                           & Paige Beukers           & paigebueckers        \\
                                           & Hanna \& Haley Cavinder & cavindertwins        \\
                                           & Khoi Young              & khoiyoung7           \\
                                           & Frederick Richards      & frederickflips       \\
                                           & Shedeur Sanders         & shedeursanders       \\
                                           & Angel Reese             & angelreese10         \\
                                           & Caitlin Clark           & caitlin.clark22      \\
                                           & Bronny James            & bronny               \\
                                           & A.J. Henning            & ajhenning            \\ \midrule
Hunting/fishing/outdoors &     BlacktipH Fishing                     & blacktiph            \\
                                           &   Ryan Izquierdo                      & ryanizfishing        \\
                                           &   jetreef                      & jetreef              \\
                                           &   RAWW Fishing                      & rawwfishingyt        \\
                                           &   kickintheirbasstv                      & kickintheirbasstv    \\
                                           &  Outdoors Weekly                       & outdoorsweekly       \\
                                           &  Frederick Penney                       & frederick            \\
                                           &  Becky Granola Girl                      & bonjourbecky         \\
                                           &   Keith Paluso                      & thisiskeithpaluso    \\
                                           &    Kween werK                       & kweenwerk    \\ \bottomrule       
\end{tabular}
}
\end{adjustbox}

%% file: aaai2026.bib
@inproceedings{diab2024classifying,
  title={Classifying Conspiratorial Narratives at Scale: False Alarms and Erroneous Connections},
  author={Diab, Ahmad and Nefriana, Rr and Lin, Yu-Ru},
  booktitle={Proceedings of the AAAI Conference on Web and Social Media (ICWSM)},
  year={2024}
}

@inproceedings{bandy2021problematic,
  title={Problematic machine behavior: A systematic literature review of algorithm audits},
  author={Bandy, Jack},
  booktitle={Proceedings of the ACM Conference on Human-Computer Interaction (CSCW)},
  year={2021},
}

@misc{paszke2019pytorchimperativestylehighperformance,
      title={PyTorch: An Imperative Style, High-Performance Deep Learning Library}, 
      author={Adam Paszke and Sam Gross and Francisco Massa and Adam Lerer and James Bradbury and Gregory Chanan and Trevor Killeen and Zeming Lin and Natalia Gimelshein and Luca Antiga and Alban Desmaison and Andreas Köpf and Edward Yang and Zach DeVito and Martin Raison and Alykhan Tejani and Sasank Chilamkurthy and Benoit Steiner and Lu Fang and Junjie Bai and Soumith Chintala},
      year={2019},
      eprint={1912.01703},
      archivePrefix={arXiv},
      primaryClass={cs.LG},
      url={https://arxiv.org/abs/1912.01703}, 
}

@inproceedings{mitra2021understanding,
  title={Understanding anti-vaccination attitudes in social media},
  author={Mitra, Tanushree and Counts, Scott and Pennebaker, James},
  booktitle={Proceedings of the AAAI Conference on Web and Social Media (ICWSM)},
  year={2021}
}

@article{jungherr2022populist,
  title={Populist supporters on Reddit: A comparison of content and behavioral patterns within publics of supporters of Donald Trump and Hillary Clinton},
  author={Jungherr, Andreas and Posegga, Oliver and An, Jisun},
  journal={Social Science Computer Review (SSCR)},
  volume={40},
  number={3},
  pages={809--830},
  year={2022}
}

@inproceedings{grover2019detecting,
  title={Detecting potential warning behaviors of ideological radicalization in an alt-right subreddit},
  author={Grover, Ted and Mark, Gloria},
  booktitle={Proceedings of the AAAI Conference on Web and Social Media (ICWSM)},
  year={2019}
}

@inproceedings{steffen2023codes,
  title={Codes, Patterns and Shapes of Contemporary Online Antisemitism and Conspiracy Narratives--an Annotation Guide and Labeled German-Language Dataset in the Context of COVID-19},
  author={Steffen, Elisabeth and Mihaljevic, Helena and Pustet, Milena and Bischoff, Nyco and Varela, Mar{\'\i}a do Mar Castro and Bayramoglu, Yener and Oghalai, Bahar},
  booktitle={Proceedings of the AAAI Conference on Web and Social Media (ICWSM)},
  year={2023},
}

@article{samory2018government,
  title={'The Government Spies Using Our Webcams' The Language of Conspiracy Theories in Online Discussions},
  author={Samory, Mattia and Mitra, Tanushree},
  journal={Proceedings of the ACM Conference on Human-Computer Interaction (CSCW)},
  year={2018}
}

@article{kou2017conspiracy,
  title={Conspiracy talk on social media: collective sensemaking during a public health crisis},
  author={Kou, Yubo and Gui, Xinning and Chen, Yunan and Pine, Kathleen},
  journal={Proceedings of the ACM Conference on Human-Computer Interaction (CSCW)},
  year={2017},
}

@inproceedings{starbird2017examining,
  title={Examining the alternative media ecosystem through the production of alternative narratives of mass shooting events on Twitter},
  author={Starbird, Kate},
  booktitle={Proceedings of the AAAI Conference on Web and Social Media (ICWSM)},
  year={2017}
}

@article{dyor_ariel,
    author = {Chinn, S. and Hasell, A.},
    title = {Support for ``doing your own research'' is associated with COVID-19 misperceptions and scientific mistrust.},
    journal = {Harvard Kennedy School Misinformation Review},
    year = {2023},
}

@article{bain23_interspeech,
  title     = {WhisperX: Time-Accurate Speech Transcription of Long-Form Audio},
  author    = {Max Bain and Jaesung Huh and Tengda Han and Andrew Zisserman},
  year      = {2023},
  journal = {INTERSPEECH 2023},
}

@article{barr2009populists,
  title={Populists, outsiders and anti-establishment politics},
  author={Barr, Robert R},
  journal={Party politics},
  volume={15},
  pages={29--48},
  year={2009},
  publisher={Sage Publications Sage UK: London, England}
}

@article{chinn2026does,
  title={What does it mean to ``do your own research?'' A comparative content analysis of DYOR messages in Instagram and Facebook posts about reproductive health, food, and vaccines},
  author={Chinn, Sedona and Hasell, Ariel and Shao, Anqi},
  journal={New Media \& Society},
  volume={28},
  pages={567--588},
  year={2026},
  publisher={Sage Publications Sage UK: London, England}
}

@article{cotter2022judging,
  title={Judging value in a time of information cacophony: Young adults, social media, and the messiness of do-it-yourself expertise},
  author={Cotter, Kelley and Thorson, Kjerstin},
  journal={The International Journal of Press/Politics},
  volume={27},
  number={3},
  pages={629--647},
  year={2022},
}

@article{eberl2021populism,
  title={From populism to the ``plandemic'': Why populists believe in COVID-19 conspiracies},
  author={Eberl, Jakob-Moritz and Huber, Robert A and Greussing, Esther},
  journal={Journal of Elections, Public Opinion and Parties},
  volume={31},
  pages={272--284},
  year={2021},
  publisher={Taylor \& Francis}
}

@article{enders2023modeling,
  title={On modeling the correlates of conspiracy thinking},
  author={Enders, Adam M and Diekman, Amanda and Klofstad, Casey and Murthi, Manohar and Verdear, Daniel and Wuchty, Stefan and Uscinski, Joseph},
  journal={Scientific Reports},
  volume={13},
  pages={8325},
  year={2023},
  publisher={Nature Publishing Group UK London}
}

@article{enders2021misinformation,
  title={Are misinformation, antiscientific claims, and conspiracy theories for political extremists?},
  author={Enders, Adam M and Uscinski, Joseph E},
  journal={Group Processes \& Intergroup Relations (GPIR)},
  volume={24},
  number={4},
  pages={583--605},
  year={2021},
}

@article{hasell2023political,
  title={The political influence of lifestyle influencers? Examining the relationship between aspirational social media use and anti-expert attitudes and beliefs},
  author={Hasell, Ariel and Chinn, Sedona},
  journal={Social Media+ Society},
  volume={9},
  pages={20563051231211945},
  year={2023},
  publisher={SAGE Publications Sage UK: London, England}
}

@article{Knight_Foundation,
  author       = {{The Knight Foundation}},
  title        = {American Views 2022: Trust, Media, and Democracy.},
  year         = {2023},
  howpublished          = {https://knightfoundation.org/reports/american-views-2023-part-2/},
 note={Accessed: 2025-05-14}
}

@article{lyons2023orientations,
  title={How orientations to expertise condition the acceptance of (mis) information},
  author={Lyons, Benjamin A},
  journal={Current Opinion in Psychology},
  volume={54},
  pages={101714},
  year={2023},
  publisher={Elsevier}
}

@article{mede2020science,
  title={Science-related populism: Conceptualizing populist demands toward science},
  author={Mede, Niels G and Sch{\"a}fer, Mike S},
  journal={Public Understanding of science},
  volume={29},
  pages={473--491},
  year={2020},
  publisher={Sage Publications Sage UK: London, England}
}

@article{merkley2020anti,
  title={Anti-intellectualism, populism, and motivated resistance to expert consensus},
  author={Merkley, Eric},
  journal={Public Opinion Quarterly},
  volume={84},
  pages={24--48},
  year={2020},
  publisher={Oxford University Press US}
}

@article{droste2021feeling,
  title={Feeling left behind by political decisionmakers: Anti-establishment sentiment in contemporary democracies},
  author={Droste, Luigi},
  journal={Politics and Governance},
  volume={9},
  pages={288--300},
  year={2021},
  publisher={PRT}
}

@article{milmo2021twitter,
  title={Twitter admits bias in algorithm for rightwing politicians and news outlets},
  author={Milmo, Dan},
  journal={The Guardian},
  year={2021},
  url={https://www.theguardian.com/technology/2021/oct/22/twitter-admits-bias-in-algorithm-for-rightwing-politicians-and-news-outlets}
}

@phdthesis{klein2023loaded,
  title={Loaded language and conspiracy theorizing},
  author={Klein, Emily},
  year={2023},
  school={Rensselaer Polytechnic Institute}
}

@article{mudde2004populist,
  title={The populist zeitgeist},
  author={Mudde, Cas},
  journal={Government and opposition},
  volume={39},
  pages={541--563},
  year={2004},
  publisher={Cambridge University Press}
}

@article{oliver2016rise,
  title={Rise of the Trumpenvolk: Populism in the 2016 Election},
  author={Oliver, J Eric and Rahn, Wendy M},
  journal={The ANNALS of the American Academy of Political and Social Science},
  volume={667},
  pages={189--206},
  year={2016},
  publisher={Sage Publications Sage CA: Los Angeles, CA}
}

@article{piazza2024populism,
  title={Populism and support for political violence in the United States: Assessing the role of grievances, distrust of political institutions, social change threat, and political illiberalism},
  author={Piazza, James A},
  journal={Political research quarterly},
  volume={77},
  pages={152--166},
  year={2024},
  publisher={SAGE Publications Sage CA: Los Angeles, CA}
}

@article{adfontesmedia,
  author       = {{Ad Fontes Media}},
  title        = {Media Bias Chart},
  year         = {2025},
  howpublished          = {\url{https://adfontesmedia.com}},
  note         = {Accessed: 2025-04-23}
}

@article{starbird2023influence,
  title={Influence and improvisation: Participatory disinformation during the 2020 US election},
  author={Starbird, Kate and DiResta, Ren{\'e}e and DeButts, Matt},
  journal={Social Media+ Society},
  volume={9},
  pages={20563051231177943},
  year={2023},
  publisher={SAGE Publications Sage UK: London, England}
}

@article{stecula2020trust,
  title={How trust in experts and media use affect acceptance of common anti-vaccination claims},
  author={Stecula, Dominik Andrzej and Kuru, Ozan and Jamieson, Kathleen Hall},
  journal={Harvard Kennedy School Misinformation Review},
  volume={1},
  number={1},
  year={2020}
}

@book{tripodi2022propagandists,
  title={The Propagandists' Playbook: How Conservative Elites Manipulate Search and Threaten Democracy},
  author={Tripodi, Francesca Bolla},
  year={2022},
  publisher={Yale University Press}
}

@article{tripodi2024your,
  title={``Do your own research'': affordance activation and disinformation spread},
  author={Tripodi, Francesca B and Garcia, Lauren C and Marwick, Alice E},
  journal={Information, Communication \& Society},
  volume={27},
  pages={1212--1228},
  year={2024},
  publisher={Taylor \& Francis}
}

@article{uscinski2021american,
  title={American politics in two dimensions: Partisan and ideological identities versus anti-establishment orientations},
  author={Uscinski, Joseph E and Enders, Adam M and Seelig, Michelle I and Klofstad, Casey A and Funchion, John R and Everett, Caleb and Wuchty, Stefan and Premaratne, Kamal and Murthi, Manohar N},
  journal={American Journal of Political Science (AJPS)},
  volume={65},
  pages={877--895},
  year={2021},
  publisher={Wiley Online Library}
}

@article{zadrozny2021carol,
  author = {Zadrozny, Brandy},
  title = {Carol's Journey: What Facebook Knew About How It Radicalized Users},
  year = {2021},
  howpublished = {\url{https://www.nbcnews.com/tech/tech-news/facebook-knew-radicalized-users-rcna3581}},
  note = {Accessed: 2025-04-23}
}

@article{armaly2024supports,
  title={Who supports political violence?},
  author={Armaly, Miles T and Enders, Adam M},
  journal={Perspectives on Politics},
  volume={22},
  pages={427--444},
  year={2024},
  publisher={Cambridge University Press}
}

@article{pew2024a,
  author       = {Colleen McClain},
  title        = {About half of TikTok users under 30 say they use it to keep up with politics, news},
  year         = {2024},
  howpublished = {\url{https://www.pewresearch.org/short-reads/2024/08/20/about-half-of-tiktok-users-under-30-say-they-use-it-to-keep-up-with-politics-news/}},
  note = {Accessed: 2024-12-14}
}

@article{pew2024b,
  author       = {Michelle Faverio},
  title        = {A majority of U.S. TikTok users are there for product reviews and recommendations},
  year         = {2024},
  howpublished = {\url{https://www.pewresearch.org/short-reads/2024/11/21/a-majority-of-us-tiktok-users-are-there-for-reviews-and-recommendations/}},
  note={Accessed: 2024-12-14}

}

@article{carrion2018you,
  title={``You need to do your research'': Vaccines, contestable science, and maternal epistemology},
  author={Carrion, Melissa L},
  journal={Public Understanding of Science (PUS)},
  volume={27},
  pages={310--324},
  year={2018},
  publisher={SAGE Publications Sage UK: London, England}
}

@article{baker2022alt,
  title={Alt. Health Influencers: how wellness culture and web culture have been weaponised to promote conspiracy theories and far-right extremism during the COVID-19 pandemic},
  author={Baker, Stephanie Alice},
  journal={European Journal of Cultural Studies (EJCS)},
  volume={25},
  pages={3--24},
  year={2022},
}

@article{chinn2023mapping,
  title={Mapping digital wellness content: implications for health, science, and political communication research},
  author={Chinn, Sedona and Hasell, Ariel and Hiaeshutter-Rice, Dan},
  journal={Journal of Quantitative Description: Digital Media},
  volume={3},
  pages={1--56},
  year={2023}
}

@article{callaghan2021feeling,
  title={Feeling good: The future of the \$1.5 trillion wellness market},
  author={Callaghan, Shaun and L{\"o}sch, Martin and Pione, Anna and Teichner, Warren},
  howpublished={\url{ https://www.mckinsey.com/industries/consumer-packaged-goods/our-insights/feeling-good-the-future-of-the-1-5-trillion-wellness-market }},
  note = {Accessed 2025-04-03},
  year={2021}
}

@article{wike2025economic,
  title={Economic Inequality Seen as Major Challenge Around the World},
  author={Wike, Richard and Fagan, Moira and Huang, Christine and Clancy, Laura and Lippert, Jordan},
  year={2025},
  howpublished={\url{https://www.pewresearch.org/wp-content/uploads/sites/20/2025/01/pg_2025.01.09_inequality_report.pdf}},
  note = {Accessed 2025-04-23}
}

@article{wellman2024friend,
  title={``A friend who knows what they’re talking about'': Extending source credibility theory to analyze the wellness influencer industry on Instagram},
  author={Wellman, Mariah L},
  journal={New Media \& Society},
  volume={26},
  number={12},
  pages={7020--7036},
  year={2024},
  publisher={SAGE Publications Sage UK: London, England}
}

@article{herrman2019tiktok,
  title={How TikTok is rewriting the world},
  author={Herrman, John},
  journal={The New York Times},
  year={2019},
  howpublished={\url{https://www.nytimes.com/2019/03/10/style/what-is-tik-tok.html}},
  note = {Accessed 2025-04-23}
}

@article{buntain2021youtube,
  title={YouTube recommendations and effects on sharing across online social platforms},
  author={Buntain, Cody and Bonneau, Richard and Nagler, Jonathan and Tucker, Joshua A},
  journal={Proceedings of the ACM Conference on Human-Computer Interaction (CSCW)},
  year={2021},
}

@article{douglas2019understanding,
  title={Understanding conspiracy theories},
  author={Douglas, Karen M and Uscinski, Joseph E and Sutton, Robbie M and Cichocka, Aleksandra and Nefes, Turkay and Ang, Chee Siang and Deravi, Farzin},
  journal={Political psychology},
  volume={40},
  pages={3--35},
  year={2019},
  publisher={Wiley Online Library}
}

@article{zadrozny2025,
  author       = {Zadrozny, Brandy},
  title        = {More than 15,000 doctors sign letter urging Senate to reject RFK Jr. as health secretary},
  year         = {2025},
  howpublished = {\url{https://www.nbcnews.com/news/us-news/rfk-jr-health-secretary-doctors-urge-senate-block-trump-administration-rcna186807}},
  note         = {Accessed: 2025-04-03}
}

@article{boyd2022development,
  title={The development and psychometric properties of LIWC-22},
  author={Boyd, Ryan L and Ashokkumar, Ashwini and Seraj, Sarah and Pennebaker, James W},
  journal={Austin, TX: University of Texas at Austin},
  volume={10},
  number={1-47},
  pages={6},
  year={2022}
}

@article{honnibal2020spacy,
  title={spaCy: Industrial-strength natural language processing in Python},
  author={Honnibal, Matthew and Montani, Ines and Van Landeghem, Sofie and Boyd, Adriane and others},
  year={2020},
  publisher={Zenodo, Honolulu, HI, USA}
}

@misc{prolific2025,
  author       = {Prolific},
  year         = {2025},
  title        = {Quickly Find Research Participants You Can Trust},
  howpublished = {\url{https://www.prolific.com}},
  note         = {Accessed 2025-01-15}
}

@misc{qualtrics2025,
  author       = {Qualtrics},
  year         = {2025},
  title        = {Qualtrics XM | The Leading Experience Management Software},
  howpublished = {\url{https://www.qualtrics.com}},
  note         = {Accessed 2025-01-15}
}

@inproceedings{bert2019,
  title={Bert: Pre-training of deep bidirectional transformers for language understanding},
  author={Devlin, Jacob and Chang, Ming-Wei and Lee, Kenton and Toutanova, Kristina},
  booktitle={Proceedings of the conference of the North American chapter of the association for computational linguistics (NAACL)},
  year={2019}
}

@article{distilbert2020,
  title={DistilBERT, a distilled version of BERT: smaller, faster, cheaper and lighter},
  author={Sanh, Victor and Debut, Lysandre and Chaumond, Julien and Wolf, Thomas},
  journal={arXiv preprint arXiv:1910.01108},
  year={2019}
}

@article{roberta2019,
  title={Roberta: A robustly optimized bert pretraining approach},
  author={Liu, Yinhan and Ott, Myle and Goyal, Naman and Du, Jingfei and Joshi, Mandar and Chen, Danqi and Levy, Omer and Lewis, Mike and Zettlemoyer, Luke and Stoyanov, Veselin},
  journal={arXiv preprint arXiv:1907.11692},
  year={2019}
}

@inproceedings{sbert2019,
  title={Sentence-bert: Sentence embeddings using siamese bert-networks},
  author={Reimers, Nils and Gurevych, Iryna},
  booktitle={Proceedings of the conference on empirical methods in natural language processing and the joint conference on natural language processing (EMNLP-IJCNLP)},
  pages={3982--3992},
  year={2019}
}

@article{lightgbm2017,
  title={Lightgbm: A highly efficient gradient boosting decision tree},
  author={Ke, Guolin and Meng, Qi and Finley, Thomas and Wang, Taifeng and Chen, Wei and Ma, Weidong and Ye, Qiwei and Liu, Tie-Yan},
  journal={Advances in Neural Information Processing Systems (NeurIPS)},
  year={2017}
}

@article{dawid1979maximum,
  title={Maximum likelihood estimation of observer error-rates using the EM algorithm},
  author={Dawid, Alexander Philip and Skene, Allan M},
  journal={Journal of the Royal Statistical Society: Series C (Applied Statistics)},
  volume={28},
  pages={20--28},
  year={1979},
  publisher={Wiley Online Library}
}

@inproceedings{hovy2013learning,
  title={Learning whom to trust with MACE},
  author={Hovy, Dirk and Berg-Kirkpatrick, Taylor and Vaswani, Ashish and Hovy, Eduard},
  booktitle={Proceedings of the conference of the North American Chapter of the Association for Computational Linguistics (NAACL)},
  year={2013}
}

@article{forbes,
   author =       "Rose, Kenny",
   title =        "Gen Z’s Social Media Dependency Is A Bridge, Not Barrier, For Advisors",
   year =         "2023",
   howpublished ={\url{https://www.forbes.com/councils/forbesfinancecouncil/2023/09/28/gen-zs-social-media-dependency-is-a-bridge-not-barrier-for-advisors/}},
   note= {Accessed: 2025-03-10}
 }

@article{fleckenstein2025representations,
  title={Representations of gender in conspiracy theories: a corpus-assisted critical discourse analysis},
  author={Fleckenstein, Kristen},
  journal={Critical Discourse Studies},
  volume={22},
  pages={357--373},
  year={2025},
  publisher={Taylor \& Francis}
}
